\documentclass[10pt,final,journal,letterpaper,twoside,twocolumn]{IEEEtran}

\usepackage{amsfonts}
\usepackage{amssymb}
\usepackage{amsmath}
\usepackage[dvips]{graphicx}
\usepackage{subfigure}
\usepackage{cite}
\usepackage{balance}
\usepackage{psfrag}

\begin{document}

\markboth{IEEE Photonics Technology Letters}{Marshoud \MakeLowercase{\textit{et al.}}: Non-Orthogonal Multiple Access for Visible Light Communications}

\title{Non-Orthogonal Multiple Access\\
for Visible Light Communications}

\author{Hanaa Marshoud,~\IEEEmembership{Student Member,~IEEE},
Vasileios~M.~Kapinas,~\IEEEmembership{Member,~IEEE},
George~K.~Karagiannidis,~\IEEEmembership{Fellow,~IEEE},
and~Sami Muhaidat,~\IEEEmembership{Senior Member,~IEEE}
\thanks{H.~Marshoud is with the Department of Electrical and Computer Engineering, Khalifa University,
Abu Dhabi, UAE (email: hanaa.marshoud@kustar.ac.ae).}%
\thanks{V.~M.~Kapinas and G.~K.~Karagiannidis are with the Department of Electrical and Computer Engineering, Khalifa University, PO Box 127788, Abu Dhabi, UAE, and with the Department of Electrical and Computer Engineering, Aristotle University of Thessaloniki, 54 124, Thessaloniki, Greece (email: \{kapinas, geokarag\}@auth.gr).}%
\thanks{S.~Muhaidat is with the Department of Electrical and Computer Engineering, Khalifa University, PO Box 127788, Abu Dhabi, UAE (email: sami.muhaidat@kustar.ac.ae), and with the Electrical and Computer Engineering Department, University of Western Ontario, Canada (email: sami.muhaidat@uwo.ca).}%
\thanks{Copyright (c) 2015 IEEE. Personal use of this material is permitted. However, permission to use this material for any other purposes must be obtained from the IEEE by sending a request to pubs-permissions@ieee.org.}%
}

\maketitle

\begin{abstract}
The main limitation of visible light communication (VLC) is the narrow modulation bandwidth, which reduces the achievable data rates. In this paper, we apply the non-orthogonal multiple access (NOMA) scheme to enhance the achievable throughput in high-rate VLC downlink networks. We first propose a novel gain ratio power allocation (GRPA) strategy that takes into account the users' channel conditions to ensure efficient and fair power allocation. Our results indicate that GRPA significantly enhances system performance compared to the static power allocation. We also study the effect of tuning the transmission angles of the light emitting diodes (LEDs) and the field of views (FOVs) of the receivers, and  demonstrate that these parameters can offer new degrees of freedom to boost NOMA performance. Simulation results reveal that NOMA is a promising multiple access scheme for the downlink of VLC networks.
\end{abstract}


\section{Introduction}

The main drawback of visible light communication (VLC) systems is the narrow modulation bandwidth of the light sources, which forms a barrier to achieving rival data rates. Recently, the development of high-rate VLC systems has been an active research area. To this end, equalization techniques \cite{equalization1}, adaptive modulation schemes \cite{adaptiveModulation1}, and multiple-input-multiple-output (MIMO) technology \cite{MIMO_VLC_Haas, MIMO_VLC_Hanaa} have been considered for achieving higher data-rates in VLC systems. Orthogonal frequency division multiplexing (OFDM) and orthogonal frequency division multiple access (OFDMA) schemes have also attracted attention in VLC systems due to their high spectral efficiency \cite{OFDM_tech1,OFDMA1}. However, conventional OFDM and OFDMA techniques cannot be directly applied to VLC systems, due to the restriction of positive and real signals imposed by intensity modulation and the illumination requirements. For this reason, DC-biasing and clipping techniques  have been proposed to adapt OFDM and OFDMA to VLC systems, but such techniques degrade the spectral efficiency and the bit error rate (BER) performance \cite{OFDMlimitation}.

\textit{Power domain multiple access}, also known as \textit{non-orthogonal multiple access} (NOMA), has been recently proposed as a promising candidate for 5G wireless networks \cite{NOMA3}. In NOMA, users are multiplexed in the power domain using superposition coding at the transmitter side and successive interference cancellation (SIC) at the receivers. In NOMA, each user can exploit the entire bandwidth for the whole time. As a result, significant enhancement in the sum rate can be achieved. Recent investigations on NOMA for RF systems are shown to yield substantial enhancement in throughput~\cite{NOMA1}.

In this paper, we propose NOMA as an efficient and flexible multiple access protocol for boosting spectral efficiency in VLC downlink (DL) systems thanks to the following reasons:
\begin{itemize}
\item NOMA is  efficient in multiplexing a few number of users \cite{NOMA1}. This is in line with VLC systems, which depend on transmitting LEDs that act as small cells to accommodate a small number of users in room environments.
\item SIC requires channel state information (CSI) at both the receivers and the transmitters to assist the fuctionalities of user demultiplexing, decoding order, and power allocation. This is a major limitation in RF but not in a VLC system, where the channel remains constant most of the time, changing only with the movement of the users.
\item NOMA performs better in high signal-to-noise ratio (SNR) scenarios \cite{NOMA1}. This is the case in VLC links, which inherently offer high SNRs due to the short  separation between the LED and the photo detector (PD), and the dominant line of sight (LOS) path.
\item VLC system performance can be optimized by tuning the transmission angles of the LEDs and the field of views (FOVs) of the PDs. These two degrees of freedom can enhance the channel gain differences among users, which is critical for the performance of NOMA.
\end{itemize}

The contribution of this paper is two-fold; first, to the best of our knowledge, this is the first work suggesting NOMA as a potential multiple access scheme for high-rate VLC systems, and second, we develop a complete framework for indoor NOMA-VLC multi-LED DL networks by adopting a novel channel-dependant power allocation strategy called \textit{gain ratio power allocation} (GRPA). GRPA significantly enhances the system's performance compared to the static power allocation approach by maximizing the users' sum rate. Moreover, the proposed framework can adjust the transmission angles of the LEDs and the FOVs of the PDs, to maximize system throughput. Notice that our framework consider users' mobility to establish a realistic scenario.

%
%

\section{System Model}\label{sec:model}
\begin{figure}
\centering%
\psfrag{L1}[][][0.7][0]{LED$_1$}%
\psfrag{L2}[][][0.7][0]{LED$_2$}%
\psfrag{f0}[][][0.7][0]{$\phi_{22}$}%
\psfrag{f1}[][][0.7][0]{$\quad\,\varphi_{1_{1/2}}$}%
\psfrag{f2}[][][0.7][0]{$\,\varphi_{11}$}%
\psfrag{f3}[][][0.7][0]{$\varphi_{13}$}%
\psfrag{f4}[][][0.7][0]{$\quad\,\varphi_{2_{1/2}}$}%
\psfrag{f5}[][][0.7][0]{$\varphi_{23}$}%
\psfrag{f6}[][][0.7][0]{$\,\varphi_{22}$}%
\psfrag{f7}[][][0.7][0]{$\phi_{11}$}%
\psfrag{f8}[][][0.7][0]{$\phi_{13}$}%
\psfrag{f9}[][][0.7][0]{$\phi_{23}$}%
\psfrag{d1}[][][0.7][0]{$d_{11}$}%
\psfrag{d2}[][][0.7][0]{$d_{13}$}%
\psfrag{d3}[][][0.7][0]{$d_{23}$}%
\psfrag{d4}[][][0.7][0]{$d_{22}$}%
\psfrag{z1}[][][0.7][0]{$z$}%
\psfrag{z2}[][][0.7][0]{$z$}%
\psfrag{K1}[][][0.7][0]{U$_1$}%
\psfrag{K2}[][][0.7][0]{U$_2$}%
\psfrag{K3}[][][0.7][0]{U$_3$}%
\includegraphics[width=1\columnwidth,trim=0 0 0 0,clip=true]{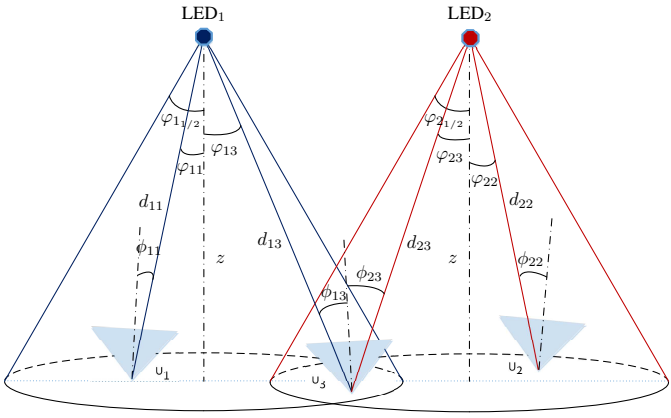}
\caption{Indoor NOMA-VLC DL system with $2$ LEDs and $3$ users.}\label{fig:VLC_network}
\end{figure}

We consider a realistic scenario with multiple LEDs in an indoor environment, such as a library or a conference room, with the beams formed by adjacent LEDs being slightly overlapped, as shown in Fig.~\ref{fig:VLC_network}. In this way, users located at the cell boundary may be receiving data streams from two adjacent LEDs. In our setup, we assume that user U$_1$ is associated or connected to LED$_1$, since it lies in the coverage of its beam. Similarly, U$_2$ is associated or connected to LED$_2$. Finally, U$_3$, located in the intersection area of the two beams, can receive data from both LEDs. Random Walk Mobility Model is implemented to mimic the movements of the indoor users. In this mobility model, a user may  move from its  location to a new location by randomly choosing a direction between $0$ and $2\pi$ and a speed between $0$ and $2$ m/s \cite{mobility_survey}.

Using NOMA, LED$_1$  transmits the real and positive signals $x_1$ and $x_3$ with power values $P_{11}$ and $P_{13}$, where $x_1$ and $x_3$ convey information intended for U$_1$ and U$_3$, respectively. Likewise, LED$_2$ transmits $x_2$ to U$_2$ and $x_3$ to U$_3$ with power values $P_{22}$ and $P_{23}$, respectively.\footnote{Therefore, U$_3$ can potentially achieve diversity gain by receiving (and combining particularly) the two copies of the same symbol $x_3$ transmitted from the two different LEDs.}\:For each LED, the transmitted signal is a superposition of the signals intended for its users. Let us now denote by $G_l$ the index set of users connected to the $l$th LED, which defines the LED$_l$ users group. Then, the signal transmitted from LED$_l$, $l=1,2$, is given~by
\begin{align}
z_l=\sum_{j\in G_l} P_{lj} x_j,
\end{align}
while the total transmitted power is $P_l = \sum_{j\in G_l} P_{lj}$.

The received signal at U$_i$ is formed by the contribution of all signals transmitted from the LEDs in the network, written~as
\begin{align}
y_i = \sum_{l=1}^L\sum_{j\in G_l} h_{li} P_{lj} x_j + n_i,
\end{align}
where $L$ is the total number of LEDs and $n_i$ denotes additive white Gaussian noise (AWGN) of zero mean and total variance $\sigma_i$, which is the sum of contributions from shot noise and thermal noise at U$_i$. The LOS path gain from LED$_l$ to U$_i$~is
\begin{align}\label{equ:channel_gain}
h_{li} = \frac{A_i}{d^2_{li}}R_o(\varphi_{li})T_s(\phi_{li})g(\phi_{li})\cos\phi_{li},\,\,0\leq\phi_{li}\leq\Phi_i,
\end{align}
while $h_{li}=0$ for $\phi_{li}>\Phi_i$. In \eqref{equ:channel_gain}, $A_i$ is the $i$th user PD area, $d_{li}$ the distance between LED$_l$ and U$_i$, $\varphi_{li}$ is the angle of irradiance with respect to the transmitter perpendicular axis, $\phi_{li}$ is the angle of incidence with respect to the receiver axis, $\Phi_i$ is the FOV of U$_i$, $T_s(\phi_{li})$ is the gain of the optical filter, and $g(\phi_{li})$ represents the gain of the optical concentrator given~by
\begin{align}\label{equ:gain}
g(\phi_{li}) = \frac{n^2}{\sin^2\Phi_i},\,\,0\leq\phi_{li}\leq\Phi_i,
\end{align}
while $g(\phi_{li})=0$ for $\phi_{li}>\Phi_i$ and $n$ is the refractive index. Also, $R_o(\varphi_{li})$ is the Lambertian radiant intensity of the LED
\begin{align}\label{equ:Lambertian}
R_o(\varphi_{li}) = \frac{m+1}{2\pi}\cos^m\varphi_{li},
\end{align}
where $m$ is the order of Lambertian emission, expressed as
\begin{align}\label{equ:m}
m = \frac{\ln 2}{\ln\cos\varphi_{l_{1/2}}}
\end{align}
with $\varphi_{l_{1/2}}$ being the transmitter semi-angle at half power.

The multi-user interference is eliminated by means of SIC and the decoding is performed in the order of increasing channel gain. Based on this order, U$_i$ can correctly decode the signals of all users with lower decoding order. The interference from users of higher decoding order (i.e., from U$_k$ with $k>i$) is not eliminated and is treated as noise. The instantaneous signal-to-interference-plus-noise ratio (SINR) for U$_i$ reads
\begin{align}\label{equ:SINR}
\gamma_i = \sum_{l=1}^L\frac{h_{li} P_{li}}{\sum_{k>i} h_{li} P_{lk} + \sigma_i^2},
\end{align}
where U$_k$ is higher than U$_i$ in the decoding order. Evidently, the strategy adopted for the allocation of the transmitted power among users is critical for the performance of the VLC system.

\section{Power Allocation for NOMA-VLC Networks}\label{sec:framework}
In this section, we present the NOMA-VLC framework with the gain ratio power allocation (GRPA) scheme. The goal is to implement NOMA in a realistic multi-LED scenario for achieving the highest possible throughput. We consider the existence of a \textit{central control unit} (CCU) that collects the necessary information about users' locations and their associated channel gains. Thanks to its deterministic nature, the VLC channel remains constant for a fixed receiver location, which simplifies channel estimation.\footnote{We can assume that an RF uplink channel is utilized for obtaining CSI.} When a user changes its location, the CCU updates its information accordingly.

\subsection{Users Association and Handover}
Users association is based on the spatial dimension \cite{last}, it exploits user position in order to determine which LED can provide access.  If the user is  located in the overlapping area of two adjacent LEDs, it will be associated with both of them. In this way, diversity gain can be achieved to enhance the performance of the cell edge users. To gain more insight, we investigate the effect of adjusting the LEDs transmission angles at half power $\varphi_{l_{1/2}}$ in order to eliminate inter-beam interference. Transmitting angle tuning can be related to cell zooming approach in \cite{cell-zooming}, where the cell size can be adjusted by modifying the transmitted power of the LED. In this case the DC component of the transmitted power needs to be calculated and changed each time the cell size needs to be altered. On the other hand, our transmitting angle tuning approach would give similar performance without the need to alter the transmitted optical power, and thus a uniform illumination can be ensured.   
For practical feasibility, we assume that each LED has two different transmission angle tunings. Once a user steps out of the coverage of its associated LED, it will be handed over to the nearest LED in order to continue its reception.

\subsection{Decoding Order}
The SIC decoding order among the users of the LED is decided based on the channel gain of each user, $h_{li}$.
By substituting \eqref{equ:gain} and \eqref{equ:Lambertian} into \eqref{equ:channel_gain} and substituting  $\cos\varphi_{li}$ with $z/d_{li}$, where $z$ is the  height between the PDs and the LEDs (which is assumed to be fixed, i.e., at table level), and assuming  vertical alignment of LEDs and PDs, the channel gain  $h_{li}$ can be expressed as
\begin{align}\label{equ:channel_simple2}
h_{li}\propto\frac{1}{d^{(m+3)}_{li} \sin^2\Phi_i}.
\end{align}
From \eqref{equ:channel_simple2}, it can be seen that the channel gain depends on two parameters; the distance and the FOV of the PD.

\subsubsection{Fixed FOVs}
If the FOVs of the PDs are fixed (i.e., not tunable), then the SIC ordering can be easily made in the decreasing order of the distance. Users in the index set of  LED$_l$ are sorted in the order of decreasing distance $d_{li}$. Then, users existing at the cell boundary (if any), are moved to the end of the decoding order. In this way, cell boundary users can decode their signals after subtracting the signals components intended for other users in both cells.
Thus, if the decreasing order of the users' distances from the LED is
\begin{equation*}
d_{lc1} > d_{lc2} > .... > d_{lcn},
\end{equation*}
for users in the cell centre, and
\begin{equation*}
d_{le1} > d_{le2} > .... > d_{len},
\end{equation*}
for cell edge users, then the decoding order is set to
\begin{equation*}
\label{equ:order}
\mho_{lc1} < \mho_{lc2} < .... < \mho_{lcn}<\mho_{le1} < \mho_{le2} < .... < \mho_{len},
\end{equation*}
where $\mho_i$ denotes the SIC decoding order for user U$_i$.

\subsubsection{Tunable FOVs}
As a further step, we examine the effect of changing the FOVs of the users. As shown in \eqref{equ:gain}, the gain of the optical concentrator of the PD can be increased by reducing the FOV. Thus, FOV tuning can be utilized to enhance channel gain differences among users, which is beneficial for the success of the NOMA technique. However, if the FOV is smaller than it should be, the PD will no longer be able to observe the required LED beam. The location of the user with respect to the transmitting LED determines the optimal FOV adjustment. We assume that each PD has three tunable FOV settings. The FOVs of cell edge users are tuned to receive the beam of one LED only (if possible) to reduce beam overlapping and enhance spectral efficiency. This is done by tuning the FOV to the lowest setting that allows reception from the nearest LED. Moreover, the FOVs of users in the center of the cell are  set to the lowest setting to enhance channel gain. Moreover, FOV tuning can be exploited to minimize the number of handovers in the system. Particularly, as the users are moving to the proximity of the transmitting LED, they can use a wider FOV setting to stay connected to the same LED and avoid unnecessary handovers. After adjusting the FOVs of the users, SIC decoding order is done based on the distance.

\subsection{Gain Ratio Power Allocation (GRPA)}
Based on the previous step, each LED  transmits the signals of its users using NOMA. To do so, different power values are allocated to the users based on their channel gains. The sum of the assigned power values is equal to the LED transmitting power. 
We propose a novel gain ratio power allocation strategy, and compare it to the static  power allocation, where the transmission power of the $i$th sorted user is set~to
\begin{align}\label{equ:fixed_power}
P_{i}=\alpha P_{i-1},
\end{align}
where  $\alpha$ is the power allocation factor $(0<\alpha<1)$. According to GRPA, power allocation depends on the user gain compared to the  gain of the first sorted user, as well as  the decoding order $i$. After  setting $i$, the  power allocated to  the $i$th sorted user~is
\begin{align}\label{equ:GRPA}
P_{i} = \left(\frac{h_{l1}}{h_{li}}\right)^i P_{i-1}.
\end{align}

Thus, the assigned power decreases with the increase of $h_{li}$, since lower power levels will be sufficient for users with good channel conditions to decode their signals, after subtracting the signals of users with lower decoding order. Moreover, the ratio $\frac{h_{l1}}{h_{li}}$ is raised to the power of the decoding order, $i$, to ensure fairness; as users with low decoding order will need much higher power due to the large interference they receive.



\section{Simulation Results and Discussion}\label{sec:results}
In this section, we evaluate the performance of the proposed NOMA-VLC framework. We consider a $6\times 6\times 3$ $\rm{m^3}$ room with two transmitting LEDs. 
In the cases with no tuning, we set the LED transmission angles and the FOVs to fixed values, i.e., $\varphi_{l} = 45$ and $\Phi_i=50$. The tunable transmission angles and FOVs are $(45,60)$ and $(15,30,50)$, respectively. We used the same LEDs and PDs characteristics as in~\cite{Komine}.

\begin{figure*}\centering%
\psfrag{xx}[t][][0.7][0]{Number of Users}%
\psfrag{yy}[][l][0.7][180]{Average Bit Error Rate}%
\psfrag{L11}[][r][0.5][0]{\, \hspace{9em} Fixed Power Alloc. ($\alpha=0.3$)}%
\psfrag{L12}[][r][0.5][0]{\, \hspace{9em} Fixed Power Alloc. ($\alpha=0.4$)}%
\psfrag{L13}[][r][0.5][0]{\, GRPA}%
\psfrag{zz}[][l][0.7][180]{Sum Rate (bps)}%
\psfrag{L21}[][r][0.5][0]{\quad\,\, No tuning}%
\psfrag{L22}[][r][0.5][0]{\quad\, $\varphi_l$ tuning}%
\psfrag{L23}[][r][0.5][0]{\quad\,\, $\Phi_i$ tuning}%
\psfrag{L31}[][r][0.5][0]{\quad\,\, No tuning}%
\psfrag{L32}[][r][0.5][0]{\quad\, $\varphi_l$ tuning}%
\psfrag{L33}[][r][0.5][0]{\quad\,\, $\Phi_i$ tuning}%
\subfigure[BER for various power allocations.]
{\includegraphics[width=0.68\columnwidth,trim=15 0 20 20,clip=true]{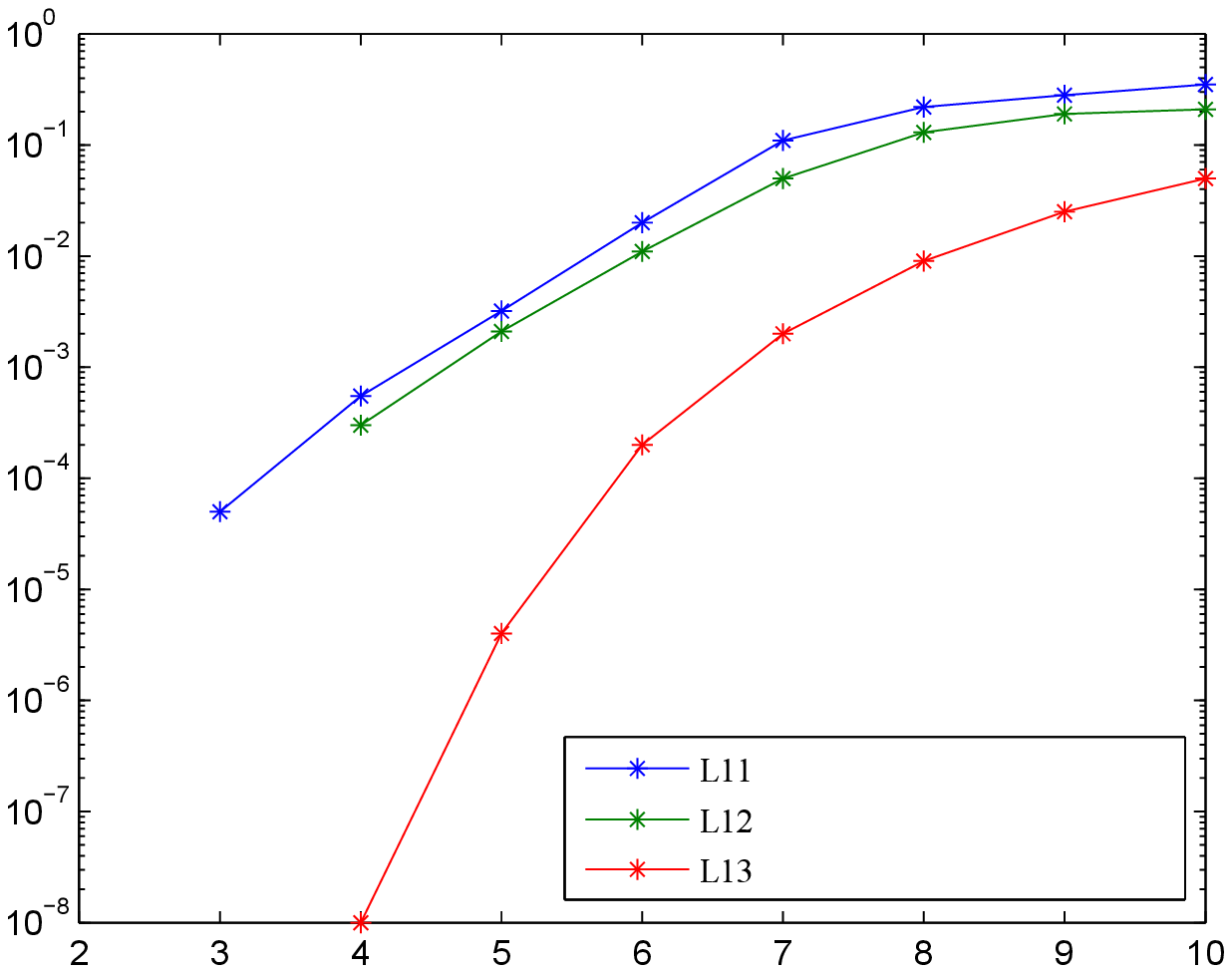}
\label{fig:fixedvsdynamic}}%
\subfigure[Sum rate with static power allocation.]
{\includegraphics[width=0.68\columnwidth,trim=15 0 20 20,clip=true]{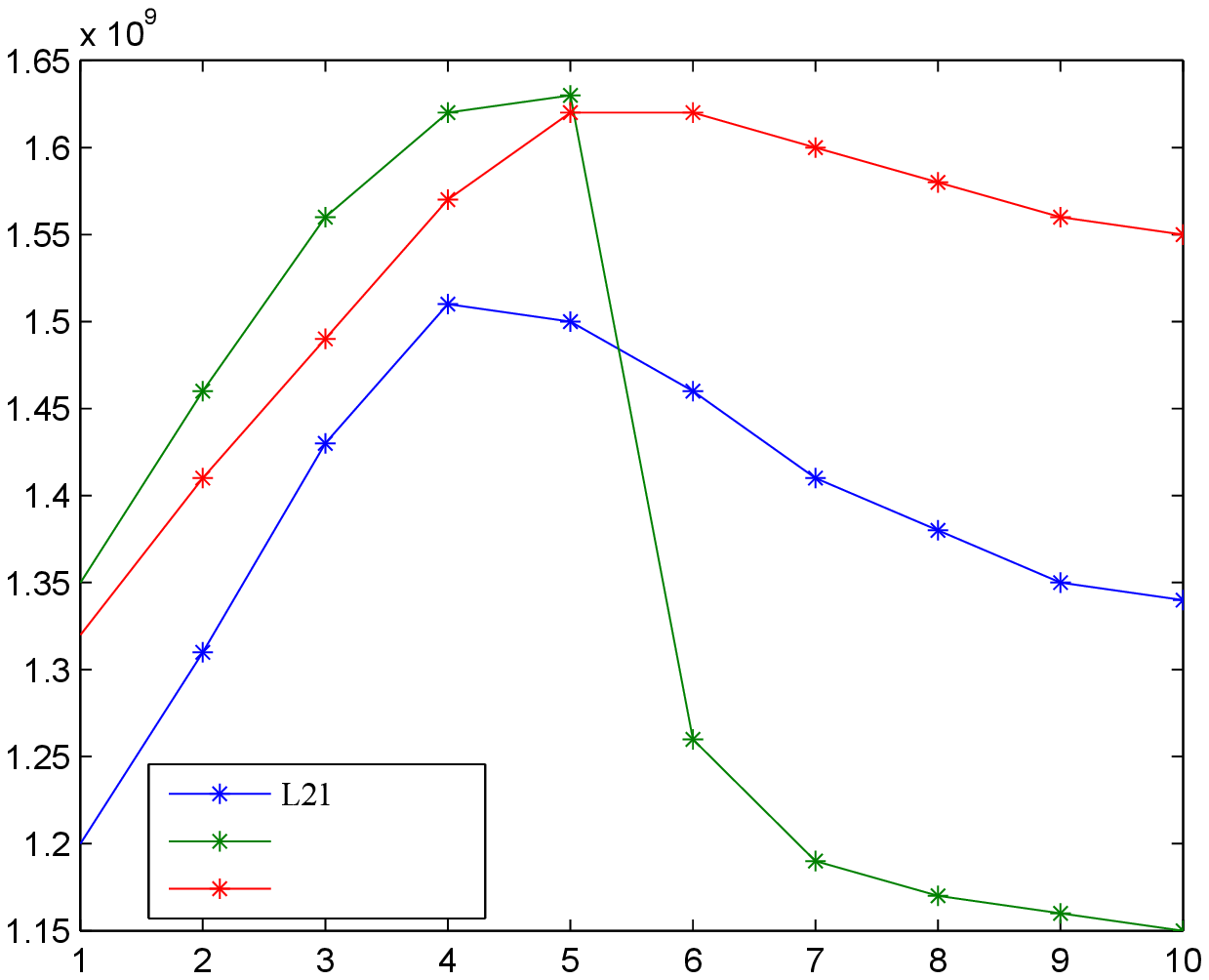}
\label{fig:SumratevsUsers}}%
\subfigure[Sum rate with GRPA.]
{\includegraphics[width=0.68\columnwidth,trim=15 0 20 20,clip=true]{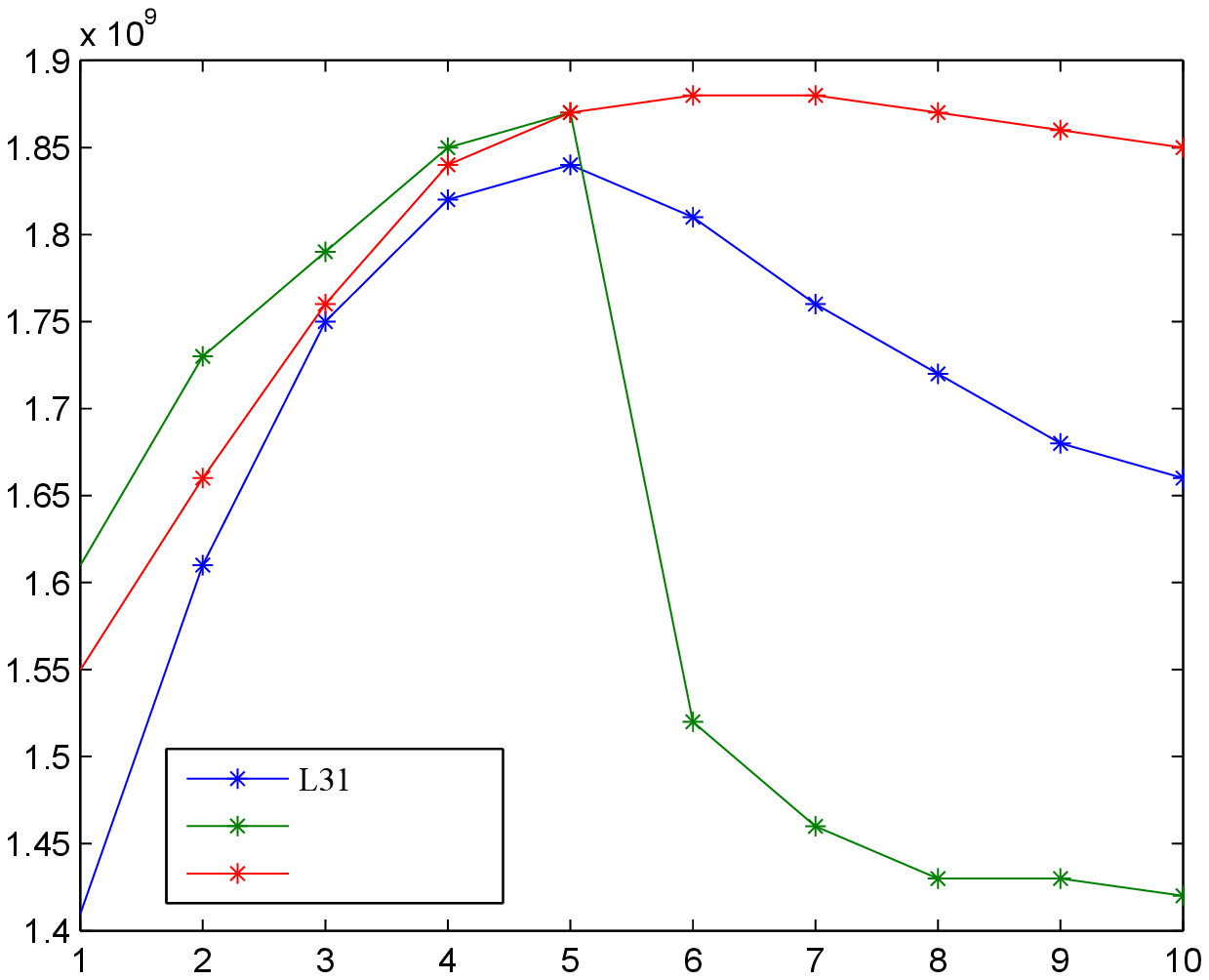}
\label{fig:SumratevsUsers_dynamic}}%
\caption{Performance of indoor ($6\times 6\times 3$ $\rm{m^3}$ room) NOMA-VLC $2$-LED DL network with respect to the number of users.}
\end{figure*}

First, we compare the BER performance for the GRPA and the static power allocation. It was found by simulations that static power allocation gives its best BER performance at ($\alpha=0.3$ and $\alpha=0.4$). At these values, the power allocated to users experiencing  bad channel conditions is high enough to enable correct signal decoding. 
Figure~\ref{fig:fixedvsdynamic} shows the BER for all users for the two power allocation schemes. The proposed GRPA strategy performs better than static power allocation as it compensates for channel differences among users. For example, at BER $=10^{-3}$, GRPA was able to serve $6$ users, while static power allocation can only accommodate $4$ users to maintain the same BER performance. It should be pointed here that GRPA is more sensitive to channel knowledge as can be inferred from \eqref{equ:GRPA}. We assumed that the RF uplink channel is noiseless in our simulations.

%
%
%

%

Next, the effect of transmission angles and FOVs tuning on system performance is studied. We primarily compare the following scenarios: 1) no tuning, 2) transmission angles tuning, and 3) FOVs tuning. Figures~\ref{fig:SumratevsUsers} and~\ref{fig:SumratevsUsers_dynamic} show the users' sum rate for the three scenarios under static power allocation ($\alpha = 0.4$) and the proposed GRPA, respectively. As it can be seen, for a small number of users, transmission angles and FOVs tuning increases the sum data rate. This is because the interference between the two beams is completely eliminated. However, as the number of users increases, less power is allocated to each user, which makes  it better for cell edge users to receive their signals from both LEDs. Thus, transmission angles tuning will degrade system throughput, as it limits cell edge users to receive from one LED only. On the other hand, FOVs tuning will have the best performance as it allows each user to optimize reception according to its position. As expected, the performance improvement induced by FOVs tuning is significantly increased when GRPA is adopted, as the latter accounts for channel gain variations. 

\begin{figure}
\centering%
\psfrag{ww}[][l][0.7][180]{Average Number of Handovers}%
\psfrag{xx}[t][][0.7][0]{Number of Users}%
\psfrag{L1}[][r][0.6][0]{\hspace{3em} Fixed FOV}%
\psfrag{L2}[][r][0.6][0]{\, \hspace{3em} Tunable FOV}%
\includegraphics[width=0.8\columnwidth,trim=30 0 30 20,clip=true]{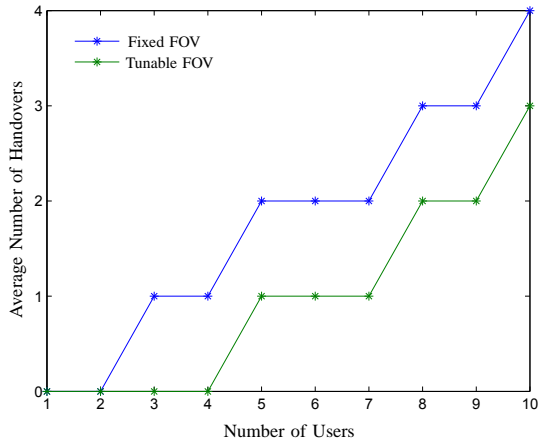}
\caption{Number of handovers in $100$ sec with respect to number of users.}\label{fig:handovers}
\end{figure}

Furthermore, FOV tuning can be exploited to decrease the number of handovers performed in the system. Figure~\ref{fig:handovers} shows the number of handovers for two scenarios: 1) fixed FOV, and 2) FOV of the moving users is adjusted to the highest setting while they move in the proximity of their associated LEDs. It can be seen that the number of handovers can be significantly decreased with the tunable FOVs strategy.

\bibliographystyle{IEEEtran}
\balance

\begin{thebibliography}{99}

\bibitem{equalization1}
H.~Li, X.~Chen, B.~Huang, D.~Tang, and H.~Chen, ``High bandwidth visible light
  communications based on a post-equalization circuit,'' \emph{IEEE Photon.
  Technol. Lett.}, vol.~26, no.~2, pp. 119--122, Jan. 2014.

\bibitem{adaptiveModulation1}
L.~Wu, Z.~Zhang, J.~Dang, and H.~Liu, ``Adaptive modulation schemes for visible
  light communications,'' \emph{J. Lightw. Technol.}, vol.~33, no.~1, pp.
  117--125, Jan. 2015.

\bibitem{MIMO_VLC_Haas}
T.~Fath and H.~Haas, ``Performance comparison of {MIMO} techniques for optical
  wireless communications in indoor environments,'' \emph{{IEEE Trans.
  Commun.}}, vol.~61, no.~2, pp. 733--742, Feb. 2013.

\bibitem{MIMO_VLC_Hanaa}
H.~Marshoud, D.~Dawoud, V.~M.~Kapinas, G.~K.~Karagiannidis, S.~Muhaidat, and B.~Sharif,
  ``MU-MIMO precoding for VLC with imperfect CSI,'' in \emph{Proc. International Workshop on Optical
  Wireless Communication (IWOW)}, Sep. 2015.

\bibitem{OFDM_tech1}
J.~Armstrong, R.~Green, and M.~Higgins, ``Comparison of three receiver designs
  for optical wireless communications using white {LEDs},'' \emph{IEEE Commun.
  Lett.}, vol.~16, no.~5, pp. 748--751, May 2012.

\bibitem{OFDMA1}
J.~Dang and Z.~Zhang, ``Comparison of optical {OFDM-IDMA} and optical {OFDMA}
  for uplink visible light communications,'' in \emph{Proc. International
  Conference on Wireless Communications Signal Processing (WCSP)}, Oct. 2012.

\bibitem{OFDMlimitation}
S.~Dimitrov, S.~Sinanovic, and H.~Haas, ``Clipping noise in {OFDM}-based
  optical wireless communication systems,'' \emph{{IEEE Trans. Commun.}},
  vol.~60, no.~4, pp. 1072--1081, Apr. 2012.

\bibitem{NOMA3}
A.~Benjebbour, Y.~Saito, Y.~Kishiyama, A.~Li, A.~Harada, and T.~Nakamura,
  ``Concept and practical considerations of non-orthogonal multiple access
  ({NOMA}) for future radio access,'' in \emph{Proc. International Symposium on
  Intelligent Signal Processing and Communications Systems (ISPACS)}, Nov.
  2013.

\bibitem{NOMA1}
Z.~Ding, Z.~Yang, P.~Fan, and H.~Poor, ``On the performance of non-orthogonal
  multiple access in $5${G} systems with randomly deployed users,'' \emph{IEEE
  Signal Process. Lett.}, vol.~21, no.~12, pp. 1501--1505, Dec. 2014.

\bibitem{mobility_survey}
T.~Camp, J.~Boleng, and V.~Davies, ``A survey of mobility models for ad hoc
  network research,'' \emph{Wireless Communications \& Mobile Computing (WCMC):
  Special Issue on Mobile Ad Hoc Networking: Research, Trends and
  Applications}, vol.~2, no.~5, pp. 483--502, Sep. 2002.

\bibitem{last}
M.~Biagi, S.~Pergoloni, and A.~Vegni, ``Last: A framework to localize, access,
  schedule, and transmit in indoor {VLC} systems,'' \emph{J. Lightw. Technol.},
  vol.~33, no.~9, pp. 1872--1887, May 2015.

\bibitem{cell-zooming}
M.~Rahaim and T.~Little, ``{SINR} analysis and cell zooming with constant
  illumination for indoor {VLC} networks,'' in \emph{Proc. International
  Workshop on Optical Wireless Communications (IWOW)}, Oct. 2013.

\bibitem{Komine}
T.~Komine and M.~Nakagawa, ``Fundamental analysis for visible-light
  communication system using {LED} lights,'' \emph{IEEE Trans. Consum.
  Electron.}, vol.~50, no.~1, pp. 100--107, Feb. 2004.

\end{thebibliography}

\end{document}